# First Results of Arar-Magnetometer Station in Saudi Arabia

Moqbil S. Alenazi, Ayman Mahrous, Ibrahim Fathy, Mohamed Nouh, Essam Elkholy, Magdy Elkhateeb, Rasha Tharwat, Heba Shalaby, Yara Ahmed

*Abstract*— **This paper presents the first results of a new Arar-magnetometer station located (Geographic Coordinates: 30°58.5′ N, 41°2.3′ E) at Northern Border University in Saudi Arabia. The geomagnetic response detected by the station during a moderate magnetic storm of April 20, 2018, is examined as an initial study. The total magnetic field components measured by the station showed a prompt increase incoincident with the Sudden Storm Commencement (SSC) time measured by ACE satellite. The high rate of magnetic field digital data system of Arar-Magnetometer station with a sampling rate of 0.1 s allowed us to study the geomagnetic pulsation at the northern region of the Arabian Peninsula for the first time.**

*Index Terms*— **Geomagnetic Pulsation, Geomagnetic Storm, Ground-Based Magnetometer**

## I. Introduction

THE satellites and ground-based magnetometers are important for making measurements of the Earth's magnetic field. They are not redundant but are instead complementary. Geomagnetic field data provide us with direct information about the magnetohydrodynamics in the Earth's outer core. Ground-based magnetometers are much less expensive and much easier to install than satellites. In addition, the contribution of ionospheric currents to the measured magnetic field by satellites should be removed; or corrected; from the field measured at the Earth's surface.

The temporal variations of the Earth's magnetic field, which is the direct or indirect effect of the solar wind interaction with the magnetosphere, are measured by Ground-based magnetometers (Mahrous, 2010). There are several types of the Ultra-Low Frequency (ULF) waves generated from the physical processes of interaction between the solar wind and Earth's magnetosphere. Geomagnetic pulsations, fluctuations, oscillations, micropulsations, hydromagnetic (HM) waves, HM emission, giant micropulsations, Rolf pulsations, pulsation trains and so on, varies names have been given to ULF waves (Gupta, 1975). Continuous (Pc) and irregular (Pi) pulsations are two broad modules of geomagnetic pulsation with short periods and small amplitudes. The magnetic field variation is a continuous function of time (Shimizu and Utada, 1999). Those of a regular and mainly continuous type is known as Pc, sub-divided into five clusters depending on their period with a range from 0.2 to 600 sec. Pulsations with an irregular and impulsive character are known as Pi and divided into two sub-groups based on a period (Jacobs et al., 1964).

The aim of this paper is to detect the geomagnetic response of the geomagnetic storm of April 20, 2018 as a first result monitored by the new Arar–magnetometer station located at Northern Border University (Geographic Coordinates Latitude: 30°58.5186′ N and 41°2.2848′ E) in Arar, the capital of Northern Borders province of Saudi Arabia. We also study the different types of geomagnetic pulsation (Pc3, Pc4, Pc5, and Pi2) detected during the geomagnetic storm. The paper is divided into four sections. The introduction is in Section 1, instrumentation and data processing are given in Section 2. In section 3, we represent the results and discussion, while in Section 4 the summary and main conclusions are given.

## II. Instrumentation and data processing

Arar-Magnetometer station is the first ground-based magnetometer station in the Arabian Peninsula, that include

Moqbil S. Alenazi, Physics Department, Faculty of Science, Northern Border University, Arar 91431-1321, Saudi Arabia. (e-mail: m.alenazi@nbu.edu.sa).
Ayman Mahrous, Space Weather Monitoring Center, Faculty of Science, Helwan University, Ain Helwan, 11795 Egypt.(e-mail: director@spaceweather.edu.eg).
Ibrahim Fathy, Basic Science Department, Egyptian Academy of Engineering and Advanced Technology, Cairo, Egypt.(e-mail: ibrahimfathy131@ eaeat.edu.eg).
Mohamed Nouh, Northern Border University, Arar, Kingdom of Saudi Arabia - NRIAG, Egypt.(e-mail: abdo_nouh@hotmail.com).
Essam Elkhouly, Northern Border University, Arar, Kingdom of Saudi Arabia - NRIAG, Egypt.(e-mail: es_ah_el@yahoo.com).
Magdy Elkhateeb, Northern Border University, Arar, Kingdom of Saudi Arabia - NRIAG, Egypt.( e-mail:  cairo_egypt10@yahoo.com).
Rasha Tharwat, Geology Department, Faculty of Science, Helwan University, Ain Helwan, 11795 Egypt.(e-mail: Rasha.Tharwat@science.helwan.edu.eg).
Heba Shalaby, Space Weather Monitoring Center, Faculty of Science, Helwan University, Ain Helwan, 11795 Egypt.(e-mail: hebashalaby@gmail.com).
Yara Ahmed, Space Weather Monitoring Center, Faculty of Science, Helwan University, Ain Helwan, 11795 Egypt.(e-mail: yara_ahmed1792@yahoo.com).

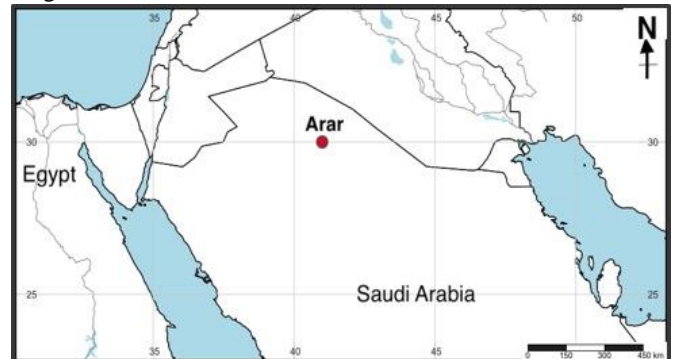

Fig. 1.  Location of Arar-Magnetometer station

the geographic area from Yemen in the south through Saudi Arabia and the Gulf States to southern Iraq in the north, which enables us to monitor the geomagnetic field over the northern region of Saudi Arabia. The location of the station is shown on the map (Fig. 1). The station's sensor is a fluxgate-type magnetometer which is considered the simplest and most common method of measuring vector magnetic fields. The magnetometer system consists of 3-axial ring-core sensors, data logging/transferring unit, power supply, and computer. The sensor house; a cubic shape room shown in Fig. 2; is connected with the control room by 2m co-axial cable. The universal time is recorded by a GPS receiver connected to the data logger/transfer unit.

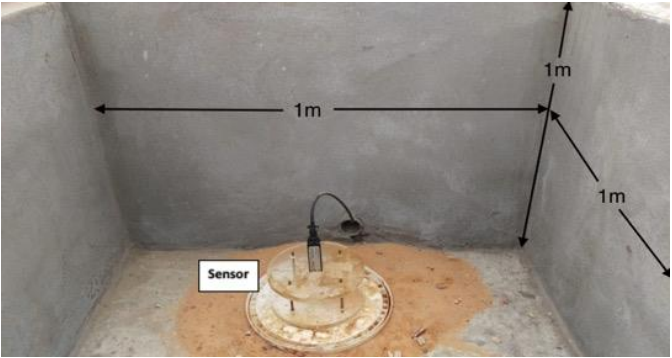

Fig. 2. Construction of the sensor house

The ambient magnetic field, expressed by horizontal (X), declination (Y), and vertical (Z) components are measured North, East and vertical downward of the geomagnetic coordinate system, respectively. These three components are digitized using field-canceling coils of dynamic range ±75,000 nT/16bit. The magnetic field digital data are obtained at the sampling rate of 0.1 s.

The X-component was filtered using zero phases shift second order butter-worth filter type according to the following equations:

$$G = \frac{(H_{oBP}) \times (ff_0)/Q}{[(f_0^2 - f^2)^2 + (ff_0/Q)^2]^{1/2}} \quad (1)$$

$$Q = \frac{f_0}{f_H - f_L}; f_0 = \sqrt{f_L f_H} \quad (2)$$

$$f_L = f_0 \left( \frac{-1}{2Q} + \sqrt{\left(\frac{1}{2Q}\right)^2 + 1} \right) \quad (3)$$

$$f_H = f_0 \left( \frac{1}{2Q} + \sqrt{\left(\frac{1}{2Q}\right)^2 + 1} \right) \quad (4)$$

Where:
G = filter gain
f0 = the filter's center frequency
Q = the quality coefficient of the filter
HoBP = the maximum gain of the filter occurring at f0

$\frac{f_0}{Q}$ = −3dB bandwidth of the filter

X-component data were filtered in the time period of range 30-40 seconds for Pc3, 70-100 seconds for Pc4, 200-500 seconds for Pc5 and 70- 100 seconds for Pi2.

### III. RESULTS AND DISCUSSION

#### A. Solar Wind Parameters

The geomagnetic storm of April 20, 2018, is considered as a moderate storm. Solar wind parameters such as speed, the density of particles and z-component of Interplanetary Magnetic Field (IMF) are measured by ACE spacecraft (http://www.srl.caltech.edu/ACE). While the ring current symmetric component and Kp index are measured by Kyoto (http://wdc.kugi.kyoto-u.ac.jp). The recorded time of the Sudden Storm Commencement (SSC) is listed in (http://isgi.unistra.fr/events_sc.php).

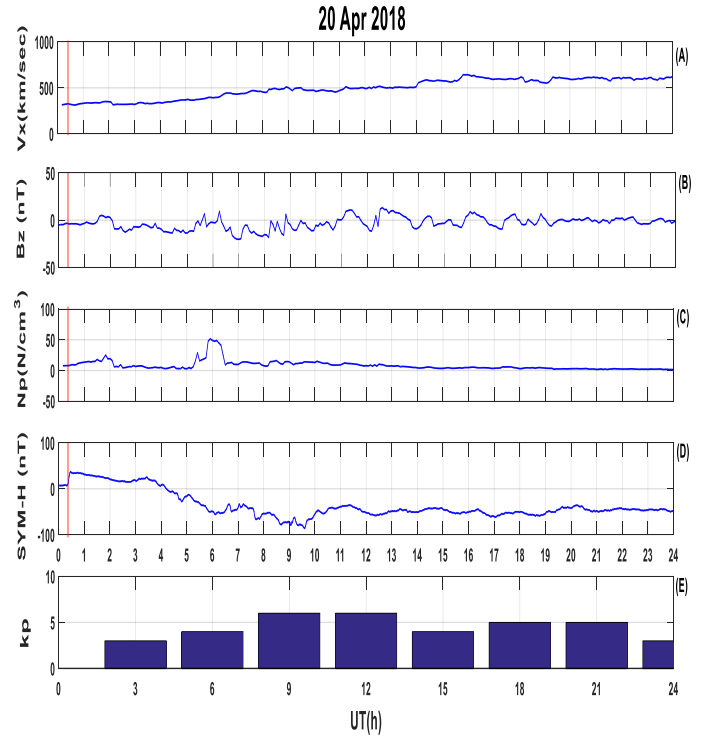

Fig. 3. Solar activity indices (A) Solar Wind speed, (B) the IMF Bz-component, (C) the particle density, (D) SYMH index (E) Kp index (three-hourly), shown versus UT for (20 April 2018) the vertical red line indicating the SSC time.

Figure 3 shows the plotting of solar wind parameters alphabetically sequenced as (A) solar wind speed, (B) IMF z-component (BZ) in Geocentric Solar Magnetospheric (GSM) coordinates, (C) the particle density, (D) the symmetric component of the ring current (SYM-H) index, (E) 3-hourly Kp index of geomagnetic activity, all versus UT during April 20, 2018 geomagnetic storm. The vertical red line in the figure indicates the SSC time. During April 20, the SSC impulse was registered at 00:21 UT, declaring the beginning of a geomagnetic storm. Solar wind speed Vx increased gradually till it reaches approximately 500 km/sec at 08:00 UT until the end of the day. The BZ value varied between (+15 to -20) nT during the whole day. The density of particles (Np) showed small variations at around 01:50 UT then increased again and reached the maximum at 06:00 UT. The SYM-H index data showed a sudden increase synchronized with the SSC time. The storm main phase continued for 9 h from the SSC, with maximum depression value (- 86 nT) at 09:30 UT. The kp index data increased gradually during the storm initial phase with value 3 and reaches maximum value 6 from 09:00 to 12:00 UT.

*B. Geomagnetic Parameters*

The results of the geomagnetic data from Arar-magnetometer station revealed high sensitivity to the geomagnetic storm of April 20, 2018. Figure (4) shows the measured geomagnetic components: (A) geomagnetic north-X, (B) geomagnetic east-Y, (C) vertical-Z, and (D) total-F, versus UT during (April 20, 2018), the red vertical line indicating the SSC time.

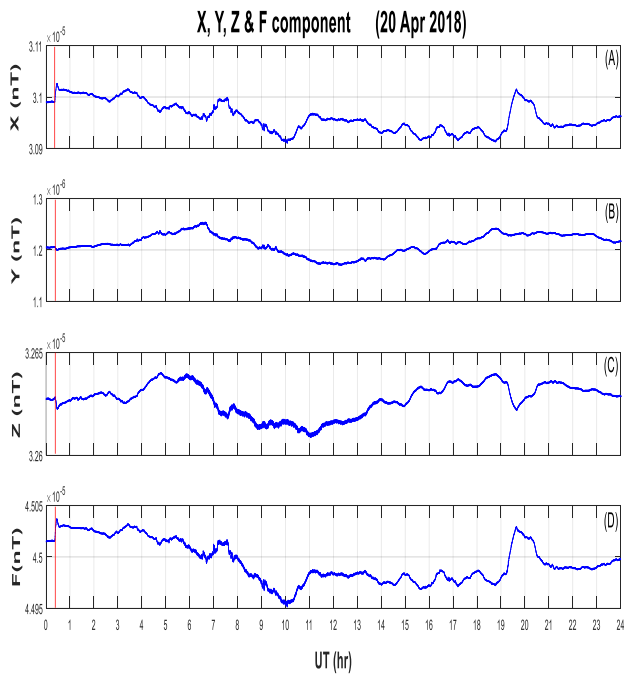

Fig. 4. Geomagnetic Components: (A) Geomagnetic North-X, (B) Geomagnetic East-Y, (C) Vertical-Z, and (D) Total-F, shown versus UT for (20 April 2018) the red vertical line indicating the SSC time.

The X-component showed a prompt increase at 00:21 UT, which is coincident with the SSC. The storm main phase continued for 9.5 h from the SSC, with maximum depression at 10:00 UT. On the other hand, the Y-component decreased slowly during 07:00~19:00 UT with a minimum value at 12:00 UT. The Z-component showed a prompt decrease at 00:21 UT, which is coincident with the SSC. Furthermore, that component showed anti-behavior of the X-component during the day. The behavior of the total intensity of the magnetic field vector (F) was almost identical to that of X-component, with the same previously mentioned features but different values of magnetic field intensity.

*C. Geomagnetic Pulsation*

The high rate of magnetic field digital data system of Arar-Magnetometer station with a sampling rate of 0.1 s allowed us to study the geomagnetic pulsation at this region. The X-component was filtered using a band-pass filter according to equations 1-4. Figure 5 shows the pulsation activity of (A) x-component, (B) Pc3, (C) Pc4, (D) Pc5 and (E) Pi2, during April 20, 2018 geomagnetic storm.

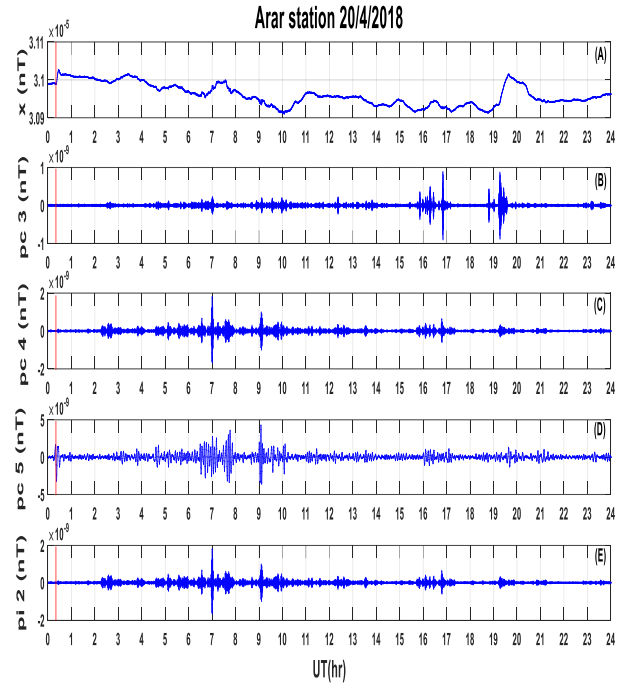

Fig. 5. Pulsation activity (A) x-component, (B) Pc3, (C) Pc4, (D) Pc5 and (E) Pi2, shown versus UT for (20 April 2018) the red vertical line indicating the SSC time.

The first type of continuous pulsations activity Pc3 (1 nt/div) showed two maxima of amplitude at 16:50 UT and 19:20 UT. On the other hand, Pc4 (2 nt/div); the second continuous pulsation type; showed the maximum amplitude at 07:00UT. It is generally accepted that some of the dayside Pc3 pulsation energy is associated with sources external to the magnetosphere. Statistical studies show that the Pc3 wave period is strongly correlated with the magnitude of the interplanetary magnetic field while the pulsation occurrence rate is dependent on the orientation of the interplanetary magnetic field (Greenstadt, E.W et al., 1980). solar wind speed and the interplanetary magnetic field are key interplanetary parameters controlling pc 3-4 pulsation activity (N. Smiljanid et al., 1998).

The behavior of Pc4 is almost identical to that of Pi2 (0.5 nT/div) except for very small amplitudes that disappeared in case of irregular pulsation. Impulsive ULF waves denoted Pi1s and Pi2s with periods of 1–40 s and 40–150 s, respectively, are observed, Pi2s observed at substorm onset are believed to be generated by the initial disturbance of the magnetospheric plasma sheet during the expansion phase onset and the propagation to the ionosphere and subsequent perturbations establish the field-aligned current (FAC) system (Olson, J. V. 1999). The last; and third; continuous pulsations activity Pc5 (5 nt/div) showed a stretching wave form resembling that of Pc4 but with different amplitude values. During periods of high solar wind speeds (450km/s), the magnetospheric cavity may become energized resulting in the excitation of monochromatic compressional ULF waves in the Pc5 band (Mann, I.R. et al., 2002).

The purpose of this study is mainly devoted to detecting the geomagnetic pulsations, Pc3, Pc4, Pi2 and Pc5 occurring on the Earth's surface which observed at Arar Magnetometer station in Saudi Arabia. Ultra low frequency (ULF) waves incident on the Earth are produced by processes in the magnetosphere and solar wind. These processes produce a wide variety of ULF hydromagnetic wave types. Waves of different frequencies and polarizations originate in different regions of the magnetosphere. The location of the projections of these regions onto the Earth depends on the solar wind dynamic pressure and magnetic field. The occurrence of various waves also depends on conditions in the solar wind and in the magnetosphere. Changes in the orientation of the interplanetary magnetic field or an increase in solar wind velocity can have dramatic effects on the type of waves seen at a particular location on the Earth. The properties of ULF waves seen at the ground contain information about the processes that generate them and the regions through which they have propagated. The properties also depend on the conductivity of the Earth underneath the observer so the study of ULF waves is a very active field of space research (McPherron, R. L., 2005)

## IV. Summary and Conclusions

The examination of the magnetic response detected by Arar-magnetometer station revealed high sensitivity to the moderate geomagnetic storm of April 20, 2018. The total magnetic field component measured by the station showed a prompt increase in the day of the storm at 00:21 UT, which is coincident with the SSC time measured by ACE satellite. The first investigation of various types of geomagnetic pulsation (Pc3, Pc4, Pc5, and Pi2) is studied during the geomagnetic storm. The distinct location of the station in the northern region of the Arabian Peninsula will support magnetospheric studies and observations in that area.

Acknowledgment

This work is supported by project no. (16-4-1436-5) funded by the Deanship of Scientific Research at the Northern Border University in Saudi Arabia. The authors are thankful to Dr. Tarek Arafa from NRIAG for his help during the deployment of Arar-Magnetometer station.